\documentclass[12pt]{iopart}
\expandafter\let\csname equation*\endcsname\relax
\expandafter\let\csname endequation*\endcsname\relax
\usepackage{amsmath}
\usepackage{hyperref}
\usepackage{xspace}
\usepackage{caption}
\usepackage{subcaption}
\usepackage{todonotes}
\usepackage[toc,page]{appendix}

\graphicspath{ {./images/} }

\def\duqtools{\emph{duqtools}\xspace}
\def\duqduq{\emph{duqduq}\xspace}
\def\imas{\emph{IMAS}\xspace}

\begin{document}

\title[Duqtools]{Duqtools: Dynamic uncertainty quantification for Tokamak reactor simulations modelling}
\author{V Azizi$^1$, S Smeets$^1$, F Koechl$^2$, F Casson$^2$, J Citrin$^{34}$\footnote{Current affiliation is Google DeepMind, London, UK}, A Ho$^3$}

\address{$^1$ Netherlands eScience Center, Science Park 402, 1098 XH Amsterdam, NL}

\address{$^2$ United Kingdom Atomic Energy Authority, Culham Science Centre, Abingdon, Oxon OX14 3DB,
United Kingdom of Great Britain and Northern Ireland}

\address{$^3$ Dutch Institute for Fundamental Energy Research, TU/e Science Park, De Zaale 20, 5612 AJ Eindhoven, NL}

\address{$^4$ Science and Technology of Nuclear Fusion Group, Eindhoven University of Technology, 5612 AJ Eindhoven, the Netherlands}
\ead{v.azizi@esciencecenter.nl}

\begin{abstract}

Large scale validation and uncertainty quantification are essential in the experimental design, control, and operations of fusion reactors. Reduced models and increasing computational power means that it is possible to run many simulations, yet setting up simulation runs remains a time-consuming and error-prone process that involves many manual steps. \duqtools is an open-source workflow tool written in Python for that addresses this bottleneck by automating the set up of new simulations. This enables uncertainty quantification and large scale validation of fusion energy modelling simulations in combination with the IMAS data model\cite{duqtools} . In this work, we demonstrate how \duqtools can be used to set up and launch 5721 different simulations of plasma experiments to validate aspects of the JINTRAC modelling suite. With this large-scale validation we identified issues in preserving data consistency in model initialization of the current ($I(p)$) distribution. Furthermore, we used \duqtools for sensitivity analysis on the QLKNN-jetexp-15D surrogate model to verify its correctness in multiple regimes.

An up to date manual of the \duqtools software can be found at \url{https://duqtools.readthedocs.io}.
\end{abstract}

\textbf{keywords:} uncertainty quantification, integrated modelling, python, jintrac, imas


\maketitle

\section{Introduction}

The next generation of magnetic fusion experiments with the tokamak configuration aims to demonstrate net energy gain for the first time, in continuous operation over long timescales \cite{bigot2022,creely:2020}. The increased complexity and scale of these experiments compared to present-day devices requires a leap in computational capabilities for predicting the behaviour of the tokamak plasma, magnets, and wall materials. This is vital for experimental design, control, and operations. Tokamak simulation suites are multi-physics, with various codes describing disparate parts of the system all interacting within a software workflow\cite{poli:2018}. Numerous such suites have been developed, such as JETTO~\cite{cenacchi:1988}, ASTRA~\cite{pereverzev:1991}, PTRANSP~\cite{transp:2018}, and ETS~\cite{kalupin:2013}. Efforts in the European fusion consortium EUROfusion, as well as at the next-generation ITER experiment, are being made to develop a modular simulation suite - the High Fidelity Plasma Simulator (HFPS). The HFPS aims to enable convenient interchange of physics modules, facilitate future development and model coupling (extensibility), be easy to maintain and deploy. This work is coupled to these ongoing efforts.


In recent years, ITER has influenced the direction of tokamak integrated modelling through the adoption of the
Integrated Modelling \& Analysis Suite (\imas) for physics modelling and analysis \cite{Imbeaux2015}. A key component of 
\imas is a standardized data model that can describe both experimental and simulation data, applicable to all
devices - not limited to ITER. The data model is open and collaboratively evolves according to the needs of
the community. To prepare for ITER operation, analysis, and supporting experiments, many research groups are making efforts to convert their tokomak data descriptions and modelling tools to the \imas data model \cite{romanelli2020}.

In this work, we present \duqtools, an open-source workflow tool written in Python for uncertainty quantification and large scale validation of fusion energy modelling simulations  in combination with the IMAS data model. One of the bottlenecks for large scale uncertainty quantification of \imas data is that setting up simulation runs involves many manual steps. This often makes it a tedious and error-prone process. \duqtools follows the same steps as one might perform for a typical sensitivity study \cite{ho2019}, but automates all the steps through workflow configuration files with minimal coding. The benefit is that workflows are standardized, completely reproducible, and can be shared with other researchers. This reduces the risk of error and makes producing confidence intervals for simulation data more accessible for non-expert, irregular, or novice users. 
\duqtools aims to work with a wide range of \imas data for fusion research. The source code is available under the terms of the Apache 2.0 licence from: \url{https://github.com/duqtools/duqtools}. An up to date manual of \duqtools can always be found at \url{https://duqtools.readthedocs.io/}.

We demonstrate two applications of \duqtools on data from plasma experiments from the Joint European Torus (JET) tokamak. We describe how \duqtools was used for a large scale validation of JINTRAC\cite{romanelli2014jintrac}, a simulation suite coupling core and edge tokamak domains and a key component of the HFPS, and a sensitivity analysis where we tested the QLKNN-jetexp-15D submodel in JINTRAC\cite{ho2021neural}.

\section{duqtools}

The goal of \duqtools is to automate the process of manipulating \imas data for (1) uncertainty quantification and (2) large scale validation. From a single template, \duqtools can set up hundreds of simulation runs. It achieves this by making use of the IMAS infrastructure through standardized interface Data Structures (IDSs) and the data directory. In this way \duqtools aims to automate workflows that researchers would normally perform manually or using ad-hoc scripts. Furthermore the automation of this process combined with the provenance of \duqtools means that the modifications performed by \duqtools are completely reproducible with minimal programming. The templates for UQ can be re-used and shared, which enables standardized sensitivity tests or canonical UQ. 

The motivation to develop \duqtools around the \imas data structure is twofold. First, standardized data from tokamak experiments worldwide facilitates simulation validation needed to improve confidence in ITER (or other machine) predictions. Second, IDSs facilitate coupling to different physics codes in simulation suites. Any software that describes its data in the \imas data structure is compatible with \duqtools.
 
In addition, \duqtools can perform batch submission of simulation runs and track their status (e.g. via Slurm or Prominence). At the end of each set of UQ runs, \duqtools can merge the simulations together to calculate simulation statistics, such as confidence ranges. 

The different steps of the \duqtools workflow are outlined in Fig \ref{fig:duqtools_flowchart}. Every \duqtools run starts with the config file, \texttt{duqtools.yaml}. Optionally, \texttt{duqtools setup} can assist in generating a config file from a shareable template. \texttt{duqtools create} takes the config file together with a template simulation and \imas data, and sets up a series of simulations for UQ. \texttt{duqtools submit} is responsible for starting or submitting jobs to a job scheduler. \texttt{duqtools status} reports the status of all currently running jobs. \texttt{duqtools merge} calculates error bounds, and \texttt{duqtools dash} launches a web-based GUI for visualizing the output data.

\duqtools can be used in multiple ways, as a command-line application, via a Python interface in scripts or notebooks, or interactively with an interactive dashboard which can be used for comparing and visualizing hundreds of simulations in one overview as confidence ranges or distributions as shown in Fig \ref{fig:dashboard}.

\subsection{Variables}

\duqtools works with a single config file (\texttt{duqtools.yaml}) in YAML format. The config file specifies data and template locations, which and how to modify the \imas data, the simulation system, and how to start the simulations. \duqtools makes the manipulation of \imas data straightforward through simple variable definitions.

Internally, these variables are defined as paths in the \imas data dictionary. For simplicity, \duqtools uses an alias to refer to the different variables. For example, the alias \texttt{t\_e} refers to \texttt{core\_profiles/profiles\_1d/*/electrons/temperature}, with dimensions \texttt{time} and \texttt{rho\_tor\_norm}. Each of these, in turn, is an alias that refers to another variable.

We distinguish between data variables, which are accessed via the \imas infrastructure as described above, and simulation parameters, which are input parameters for, e.g., JETTO. These are written to a text file through a tool like \texttt{jetto-python-tools}. These could be machine parameters, such as the reference major radius ($R_0$) or the start/end time of the simulation.

Although \duqtools has a default set of variables and aliases, users can configure and add their own variables, aliases, and dimensions. 

\paragraph{Available methods for working with variables:}

\begin{itemize}
    \item \textbf{Hypercube Sampling}: 
      Hypercube sampling is at the core of the modifications that \duqtools makes to input \imas data. 
      Dimensions for the hypercube are specified through the config file, where the \texttt{variables}(\texttt{dimensions}) and \texttt{values} can be specified to sample from.
    \item \textbf{Operations}: Modifications to the \imas data re specified via operations. These can be updates to the machine parameters, simulation metadata, data fixes that must always be applied to the data. For example, this can be used to update the major radius in the template to a different value.
    \item \textbf{Variable specification}: To work with \imas data, \duqtools must know how to navigate the \imas data structure. \duqtools includes an internal variable look-up table (which can also be user-specified) that maps variable names to their location in the \imas data directory. This enables short names that are easier to remember and specify. In the example below, \texttt{t\_e} and \texttt{zeff} map to $t_e$ and $Z_{\mathrm{effective}}$ in the core profiles IDS, respectively, with time and $\rho_{\mathrm{tor,norm}}$ as dimensions.
    \item \textbf{Coupled variables}: 
        \duqtools also supports coupling of variables. This can be used to linearly correlated variables in operations (that are mainly used for sampling). 
\end{itemize}

\subsection{Modes of operation}

\duqtools was designed for uncertainty quantification (UQ). \duqtools can be used to perform UQ on a single run, or automate the process for multiple runs. We recognized early on that configurations must be sharable by users to perform standard procedures or canonical UQ. These are enabled through template-based run generation. The different modes of operation are outlined below:

\begin{itemize}
    \item \textbf{Single run}: The default mode of operation for \duqtools is to start with a single simulation (e.g. JETTO) template, and \imas data set from a single config. \duqtools can be used to apply the variable operations to the data set and sample from the hypercube along the dimensions specified. \duqtools can handle job submission and status management. The resulting data sets can be aggregated automatically for UQ.
    \item \textbf{Templated run}: During the development we found that a single config is not sufficient if we want to analyse multiple data sets. For example, when using multiple data sets with different start and end times. \texttt{duqtools setup} can read these variables from the data set directly and write to the required fields in the \duqtools config on-the-fly.

    The advantage of template-based runs is that the UQ run can be pre-configured. This means the templates can be shared between users. The user specifies the location of the template \imas data, the name of the run, and the output directly.
    \item \textbf{Multiple runs}: For working with hundreds or even thousands of data sets, we developed an additional tool called \duqduq. This is a tool that automates the setup, creation, submission, and aggregation of large numbers of UQ runs. This tool mimics the steps outlined in Fig.~\ref{fig:duqtools_flowchart}. The difference is that the \imas handles are provided through a \emph{csv} file, and \duqduq loops over the data sets while performing the various operations. This enables large scale validation of \imas data and tokomak simulations from a single \duqtools template.
\end{itemize}

\subsection{Data aggregation and analysis}

After running experiments with \duqtools the user is usually left with a large number of output data. \duqtools provides various functions to handle and analyse this data, provided that it is generated in the \imas format. We have extracted the functionality to convert \imas data to xarray in a separate standalone tool called imas2xarray\cite{imas2xarray}. This tool can be used by scientists if no other functionality than converting \imas to xarray format is required.

\begin{figure}
    \centering
    \begin{subfigure}[ht!]{0.475\textwidth}
    \includegraphics[width=\textwidth]{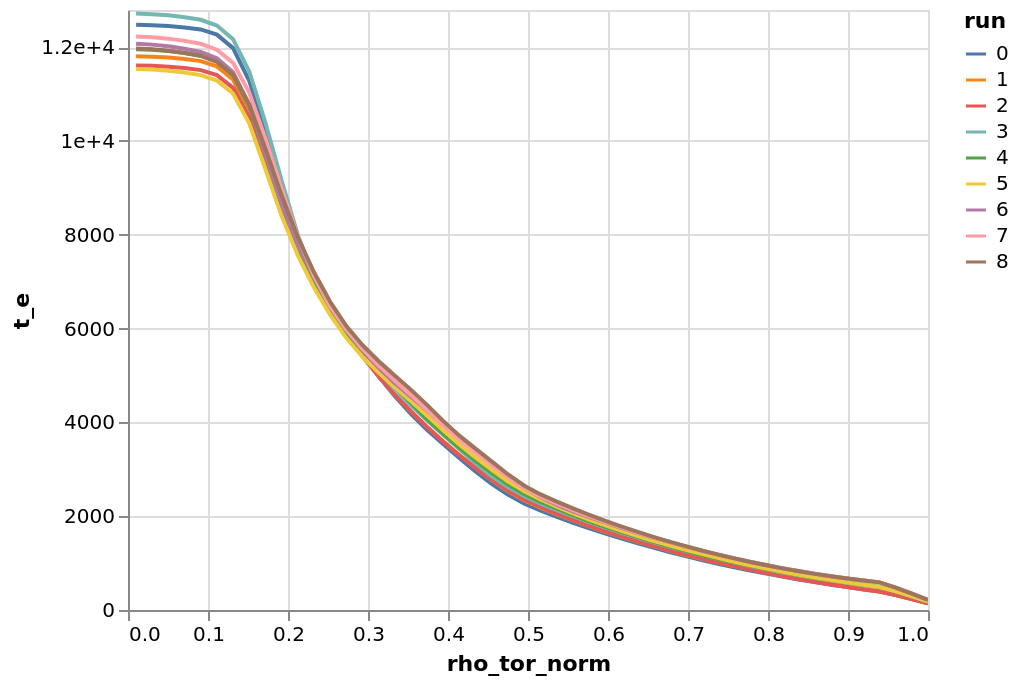}
    \caption{Plot generated by \duqtools for a local sensitivity test of $Z_{eff}$ and $T_e$ for a single predictive simulation.}
    \label{fig:plot-t-e}
    \end{subfigure}%
    \hspace{0.04\textwidth}%
    \begin{subfigure}[ht!]{0.475\textwidth}
    \includegraphics[width=\textwidth]{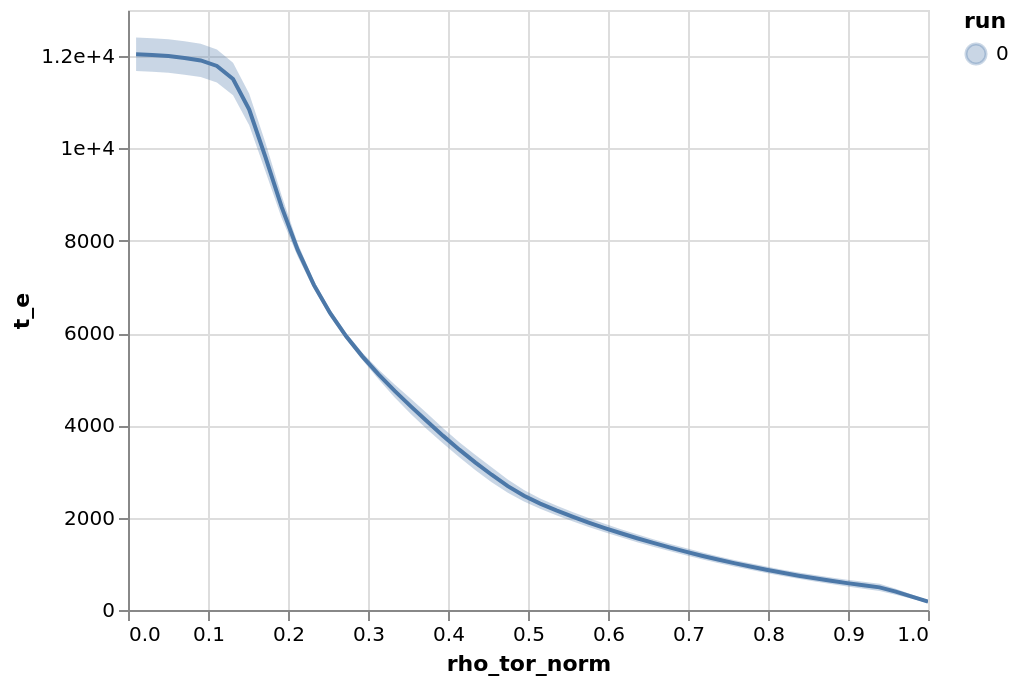}
    \caption{Same plot as (a), Except output runs are merged beforehand and the data is plotted with the standard deviation}
    \label{fig:plot-t-e-merged}
    \end{subfigure}  
    \caption{}
\end{figure}

After UQ runs the user will have many output \imas files. To aid in analyzing the data \duqtools has the capability to Summarize the statistics of multiple \imas output files from UQ runs to a single \imas file with error bars and the mean filled in, assuming Gaussian statistics (Fig.~\ref{fig:plot-t-e-merged}). \duqtools can aggregate the data, even if the data has different resolutions by using interpolatiohttps://cloud.lipsum.eu/remote.php/webdav/ns over any dimension with the help of the xarray library\cite{hoyer2017xarray}. 

\duqtools provides functionality for creating interactive plots from any \imas data, by using the \emph{altair} package \cite{VanderPlas2018}. \duqtools has built in support for most commonly used \imas dimensions, and has support for plotting error bars present in \imas data as well (Fig.~\ref{fig:plot-t-e}).

\begin{figure}
    \centering
    \begin{subfigure}[b]{0.475\textwidth}
    \includegraphics[trim=0 0 115 0 ,width=0.40\linewidth]{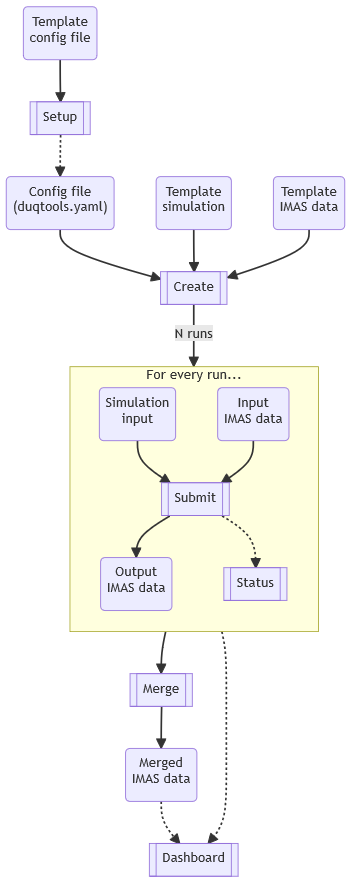}
    \caption{Flowchart showing the typical \duqtools UQ run. }
    \label{fig:duqtools_flowchart}
    \end{subfigure}
    \begin{subfigure}[b]{0.5\linewidth}
    \centering
    \includegraphics[width=0.9\textwidth]{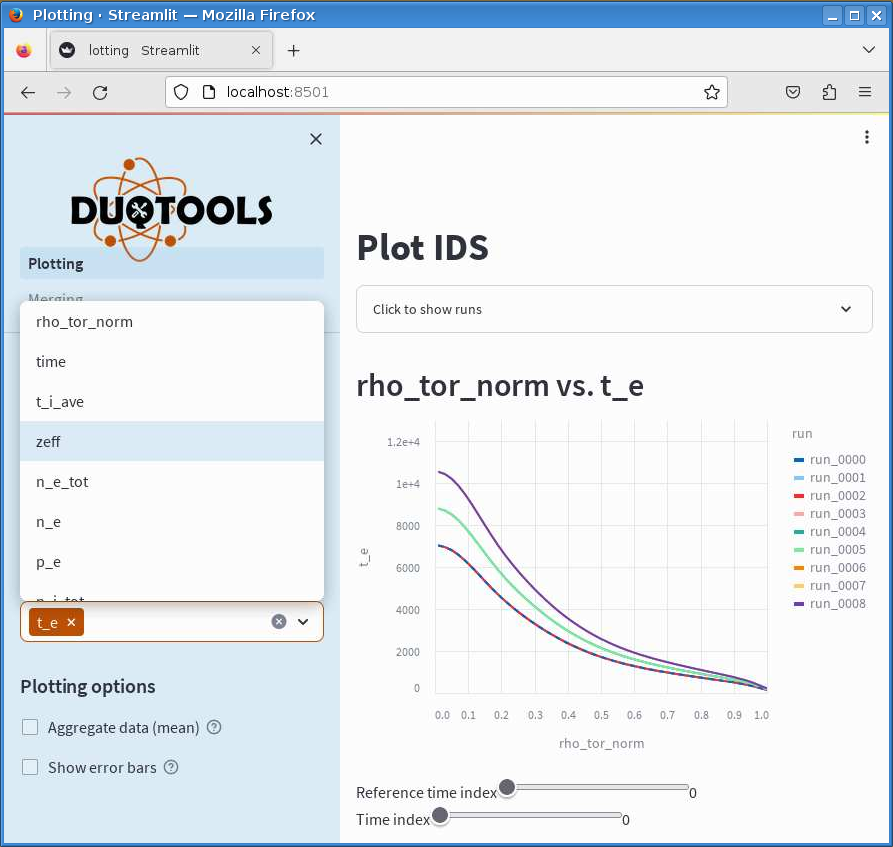}
    \caption{Example of how to create plots in the dashboard}
    \label{fig:dashboard}
    \end{subfigure}
    \caption{}
\end{figure}

Finally, \duqtools includes an interactive dashboard that is meant to explore the data from multiple runs (Fig.~\ref{fig:dashboard}).

\clearpage

\section{Applications}

\subsection{Large scale validation}

Using the automation provided by \duqtools significantly reduces the workload barrier to scientific studies requiring the large-scale execution of the JINTRAC\cite{romanelli2014jintrac} model. As an example, we used \duqtools in conjunction with an automated plasma state extraction tool~\cite{ho2019}, to setup and launch 5721 individual simulations. Each of these simulations corresponds to inputs taken from a 0.5~s steady-state time window from across 2000 different plasma experiments from the Joint European Torus (JET) tokamak in Culham, UK \cite{ho:2021}. The \duqtools configuration files used for this experiment have been made available on zenodo \cite{ho_2024_11120000} 

The simulations were run in interpretive mode, meaning that none of the principal magnetic or kinetic plasma quantities were updated from their input values during the simulation. Thus, only the derived quantities should have changed in the simulation output, effectively representing their corresponding values when computed directly from the principal input plasma state. This exercise serves as a sanity check that the extracted plasma representation used as input is self-consistent and that the processed plasma parameters computed internally within the code suite using this representation are similar to those inside the experimental database.

\begin{figure}
    \centering
    \begin{subfigure}[ht!]{0.475\textwidth}
    \includegraphics[width=\textwidth]{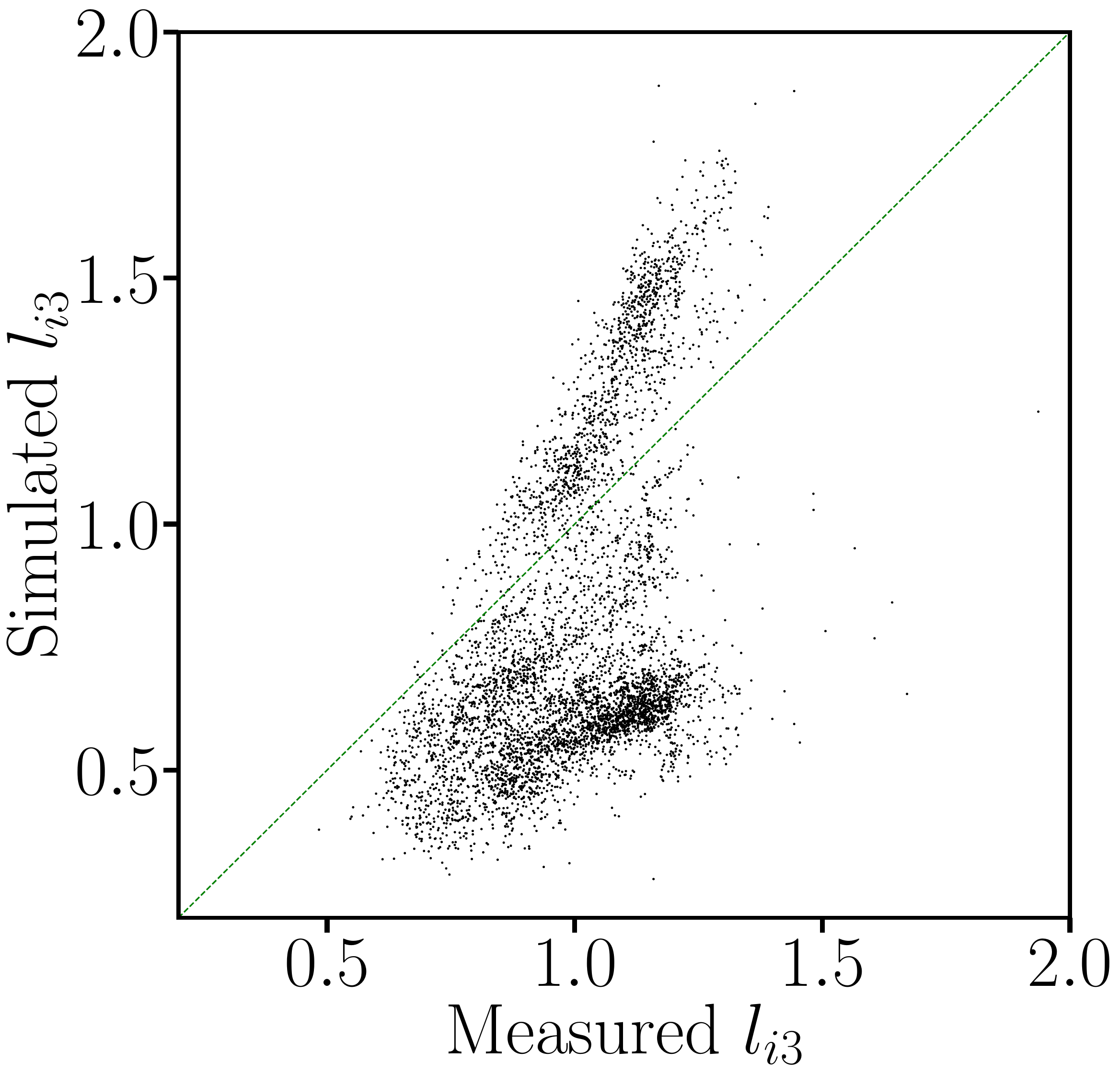}
    \end{subfigure}%
    \hspace{0.04\textwidth}%
    \begin{subfigure}[ht!]{0.445\textwidth}
    \includegraphics[width=\textwidth]{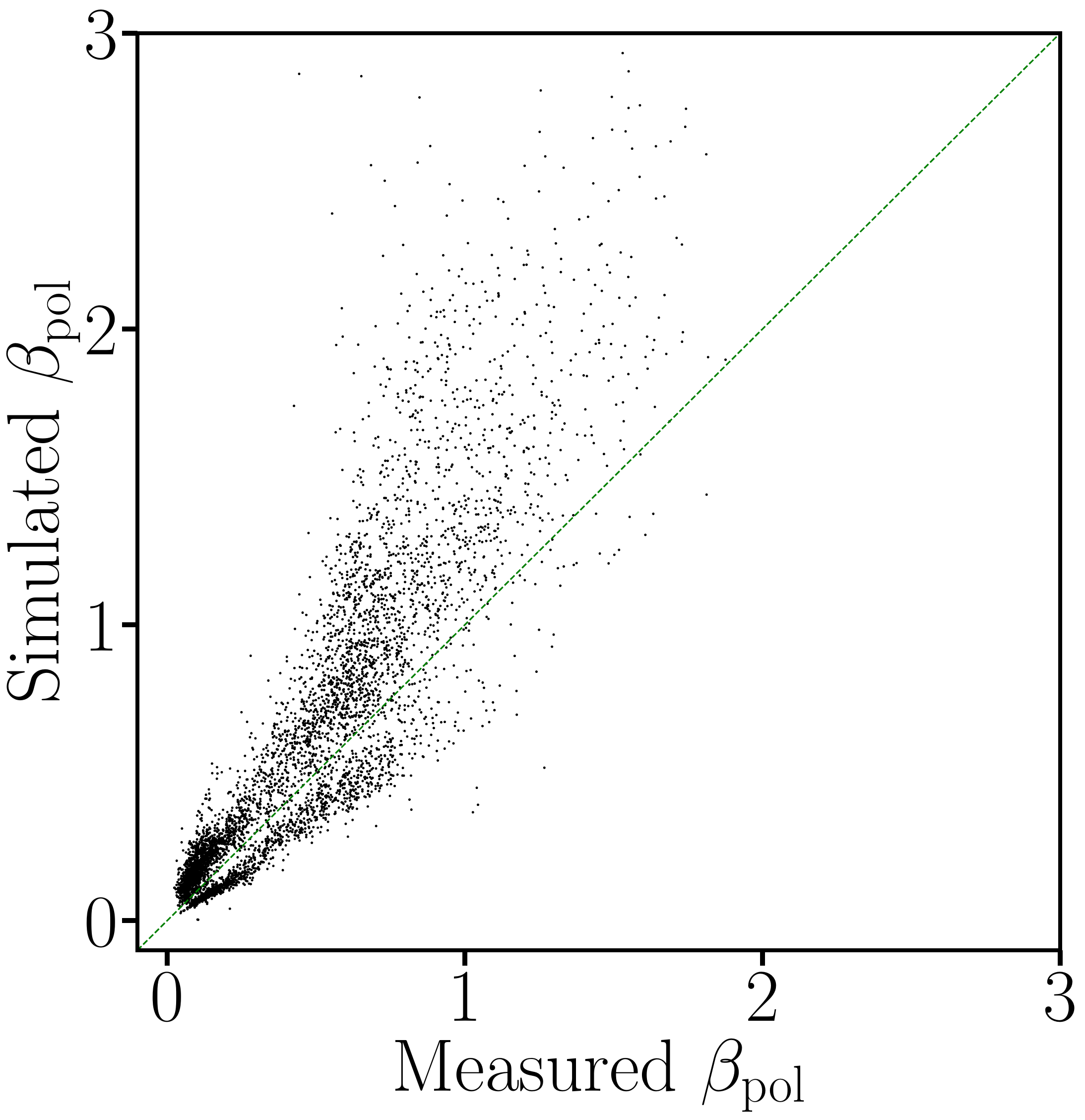}
    \end{subfigure}
    \caption{Comparison between input (Experimentally inferred) and output (processed) values for the normalized plasma internal inductance, $l_{i3}$, and the normalized poloidal magnetic field, $\beta_{\text{pol}}$, using simulation inputs processed from the experimental database prior to large-scale analysis.}
    \label{fig:large_scale_before_fix}
\end{figure}

Figure~\ref{fig:large_scale_before_fix} shows the comparison between input (Experimentally inferred)\cite{szepesi2021advanced} and output (processed) values of two derived quantities, the normalized plasma internal inductance, $l_{i3}$, and the normalized volume-averaged poloidal magnetic field, $\beta_{\text{pol}}$, evaluated at the last-closed-flux-surface. These particular parameters are important as they represent quantities used within magnetic control models stabilize the vertical position of the experimental plasma. As shown, a significant scatter was found in these parameters leading to concerns over both the data extraction routine and the internal data representation.

The source of this discrepancy was identified to be as follows. In this simulation scan, the JINTRAC input safety factor profile $q\!\left(\rho\right)$ was obtained from the EFIT equilibrium code. However, there was no constraint to ensure that the mapping of $q$ from EFIT to JINTRAC maintains the total plasma current $I_p$ at the edge. To maintain consistency, the JINTRAC code was internally scaling the entire input safety factor profile on simulation initialization. However, this rescaling maintained large spurious current densities towards the plasma edge which arose as a result from the mapping artifacts, impacting the JINTRAC representation of the current density profile and thus poloidal magnetic field, which in turn modifies the result from the calculation of the $l_{i3}$ and $\beta_{\text{pol}}$ parameters. 

Although this rescaling operation is necessary to ensure consistency within the internal representation of the plasma state, its chosen implementation clearly disrupts its consistency with the expected input plasma state. To fix the issue, we adjusted the modification such that it only applied to the outer region of the $q$ profile, where the mapping issues arise. In this way the value at $\rho = 0.95$ is left constant as it is typically used to identify the plasma scenario. 
We note that in interpretive simulations where current diffusion is allowed to self-consistently evolve, the rescaling operation impacts the initial state, but the impact of the artifacts are reduced in subsequent evolution.

\begin{figure}
    \centering
    \begin{subfigure}[ht!]{0.475\textwidth}
    \includegraphics[width=\textwidth]{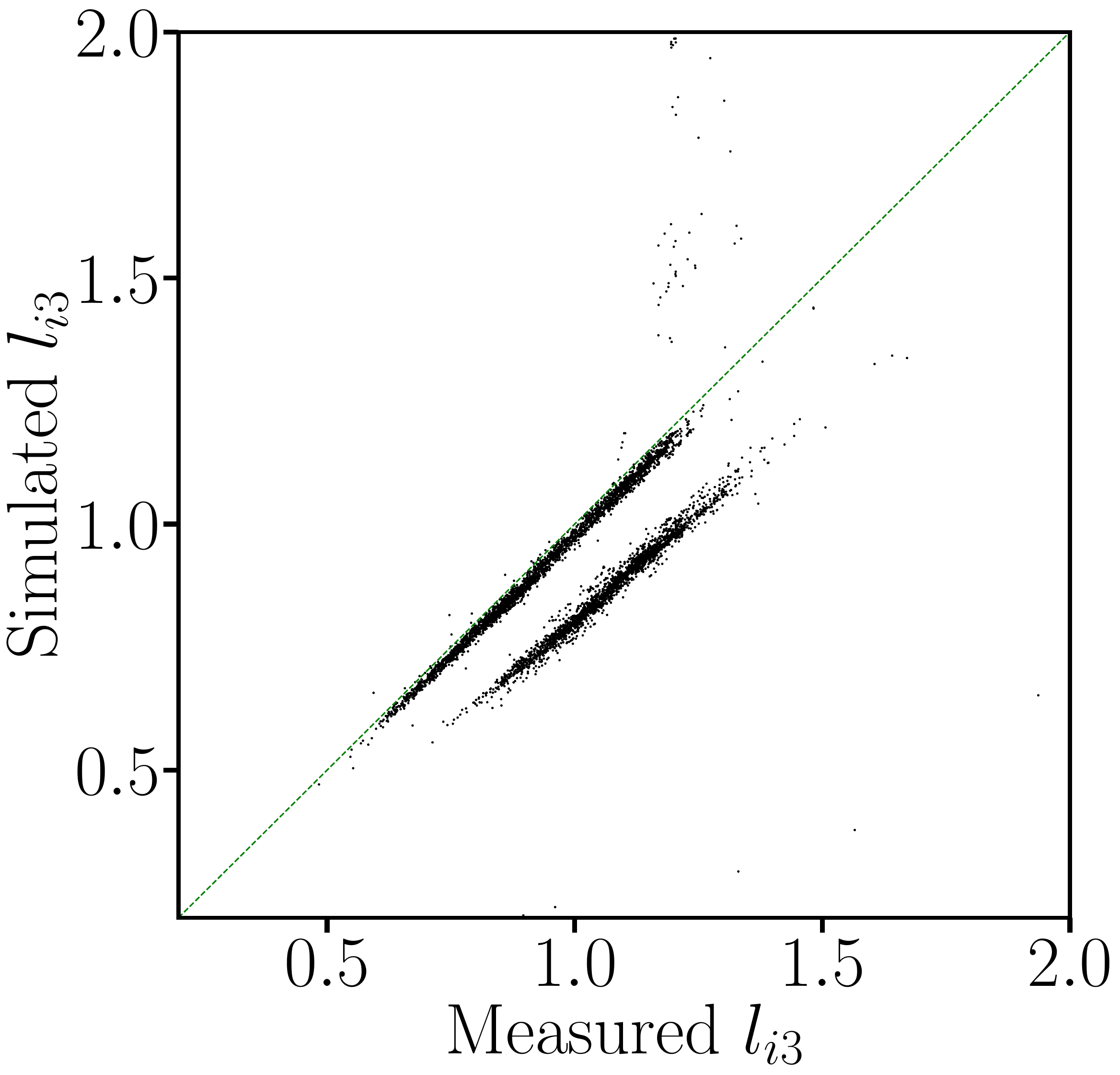}
    \end{subfigure}%
    \hspace{0.04\textwidth}%
    \begin{subfigure}[ht!]{0.445\textwidth}
    \includegraphics[width=\textwidth]{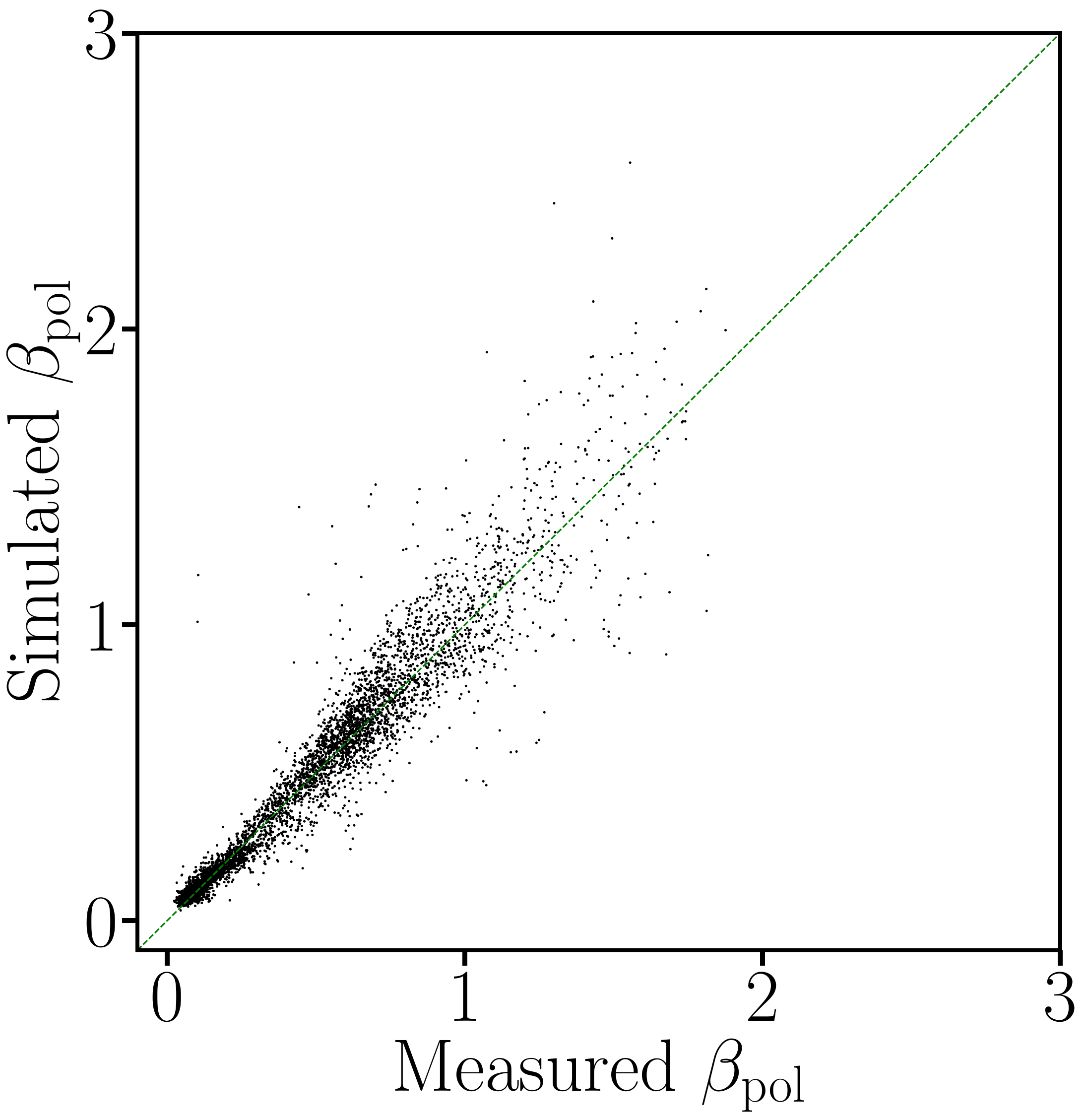}
    \end{subfigure}
    \caption{Comparison between input (measured) and output (simulated) values for the normalized plasma internal inductance, $l_{i3}$, and the normalized poloidal magnetic field, $\beta_{\text{pol}}$, using simulation inputs processed from the experimental database after corrections discovered using large-scale analysis.}
    \label{fig:large_scale_after_fix}
\end{figure}

Figure~\ref{fig:large_scale_after_fix} shows the comparison of these two parameters after applying the aforementioned fix. This reveals two distinct clusters in the internal inductance comparison. Further analysis revealed that this is because of an error in the extraction routine which used an alternate measurement signal for the derived quantity, effectively introducing a systematic shift in the measured value.

In the light of this rudimentary large-scale analysis, the execution and analysis of simulation swarms such as this allows discrepancies to be identified with a lower chance of selection bias and with greater clarity in determining the source of the discrepancy. The automated capabilities provided by \duqtools allows the setup of all these runs to be completed in approximately 4 hours, limited mostly by file read/write operations on the distributed computing system and significantly accelerated through the use of the \imas HDF5 backend. This marks a dramatic reduction of human time compared to the existing manual setup method, which was previously performed via a graphic user interface (GUI).

\subsection{Sensitivity studies}

While the automation of transferring metadata from \imas-formatted data to the simulation inputs significantly lowers the entry barrier to working with the JINTRAC simulation suite, most physics-related analysis involves setting up multiple simulations from the same reference case with minor variations between them. Thus, as a further example of versatility offered by \duqtools, this tool was used in a sensitivity study of a single time window from the set of 5721 discussed in the previous section. As with the previous example, the chosen simulation input corresponds to a 0.5~s steady-state time window extracted from an experiment from the JET tokamak.

The simulations were run in predictive mode, meaning that a selection of the magnetic and/or kinetic plasma quantities were updated from their input values according to the underlying set of 1D transport equations during the simulation. For the specific settings chosen for this study, the current, electron heat, ion heat, and ion particle transport equations were actively evaluated and their corresponding quantities evolved in time. In these simulations, ESCO was used as the fixed-boundary equilibrium solver, SANCO as the impurity transport solver and line radiation model, NCLASS\cite{houlberg:1997} as the neoclassical transport model, and QLKNN-jetexp-15D\cite{ho:2021} as the fast turbulent transport model. An ad-hoc additional diffusive transport term was applied dynamically to the region where $q < 1.1$, in order to emulate the effect of sawtooth instabilities in a continuous manner. As the inputs were taken from steady-state time windows, the simulated plasma time was chosen to be approximately long enough in order for the plasma to reach a pseudo-steady-state. This allows for a reasonable comparison between the last time slice of the simulation and the input profiles representing the experimental plasma.

As there is some uncertainty in the exact value of these profiles, a sensitivity of the simulation result to the exact value of the input profiles within this uncertainty is often performed. Thus, 5 different input 1D profiles quantities relevant to the JINTRAC simulation suite were simultaneously varied for a single discharge. These quantities are the electron temperature, $T_e$, the ion temperature, $T_i$, the electron density, $n_e$, the effective charge, $Z_{\text{eff}}$, and the safety factor, $q$. To keep the exercise simple, the variations of the profiles were limited in value to discrete multiples of the combined measurement and fit uncertainties of said input profiles, as extracted from a standard fitting routine used for JET experimental data~\cite{ho2019}. Table~\ref{tbl:input-hypercube-for-sampling} shows the specific values of these multipliers per input variable, which were then down-sampled using the Latin-hypercube sampling method to limit the computational time required to gain valuable insight.

\begin{table}
    \centering
    \begin{tabular}{c|c}
    Variable & Values \\
    \hline
    $T_e$ & $\left[-2, -1, 0, 1, 2\right] \cdot \sigma$\\
    $T_i$ & $\left[-2, -1, 0, 1, 2\right] \cdot \sigma$\\
    $n_e$ & $\left[-2, -1, 0, 1, 2\right] \cdot \sigma$\\
    $Z_{\text{eff}}$ & $\left[-1, -0.5, 0, 0.5, 1\right] \cdot \sigma$\\
    $q$ & $\left[-2, -1, 0, 1, 2\right] \cdot \sigma$
    \end{tabular}
    \caption{Input profile variation values used to form the hypercube which was sampled within the sensitivity study.}
    \label{tbl:input-hypercube-for-sampling}
\end{table}

\begin{figure}
    \centering
    \begin{subfigure}[ht!]{0.45\textwidth}
    \includegraphics[width=\textwidth,trim=0 0 130 0,clip]{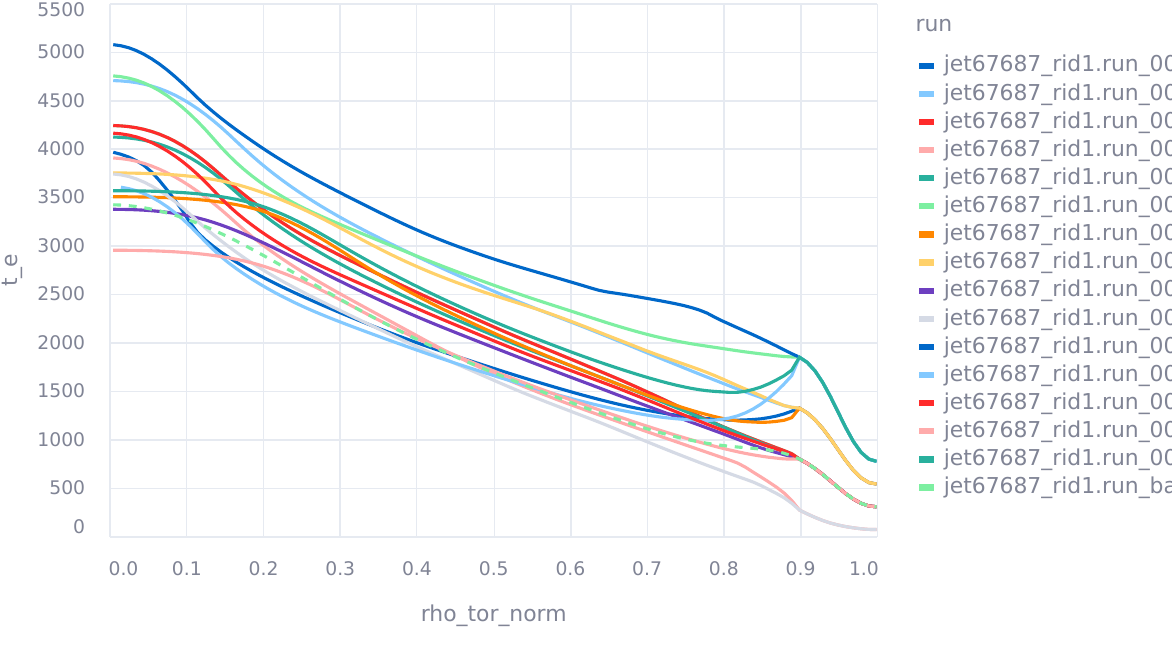}
    \end{subfigure}%
    \hspace{0.02\textwidth}%
    \begin{subfigure}[ht!]{0.45\textwidth}
    \includegraphics[width=\textwidth,trim=0 0 130 0,clip]{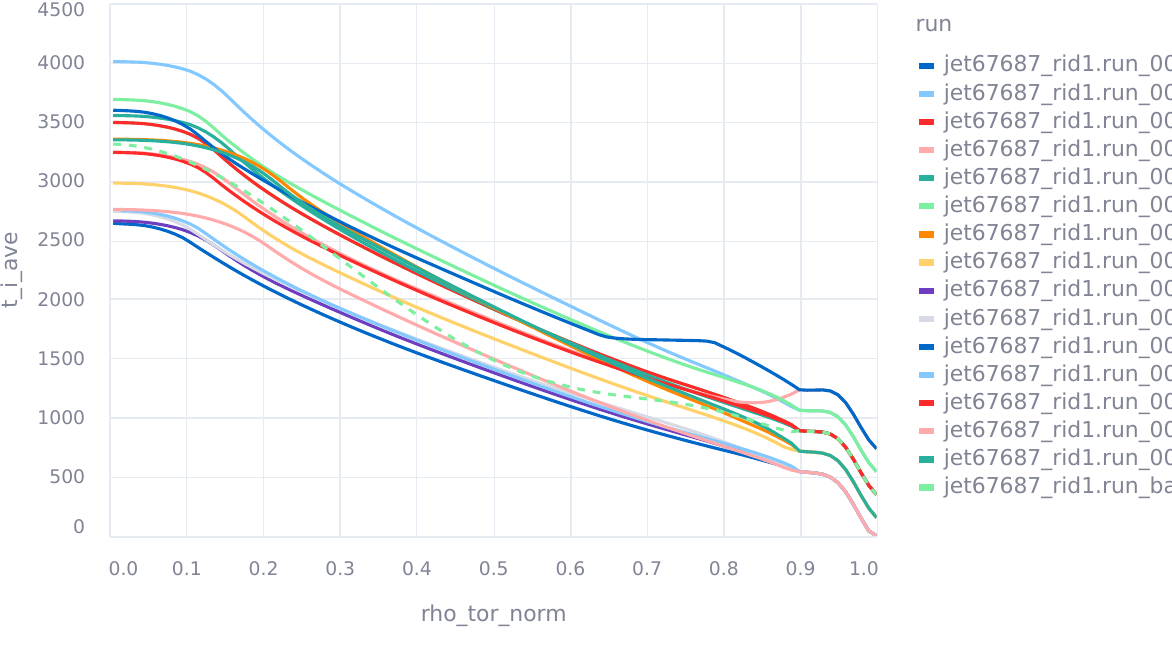}
    \end{subfigure}\\
    \begin{subfigure}[ht!]{0.45\textwidth}
    \includegraphics[width=\textwidth,trim=0 0 130 0,clip]{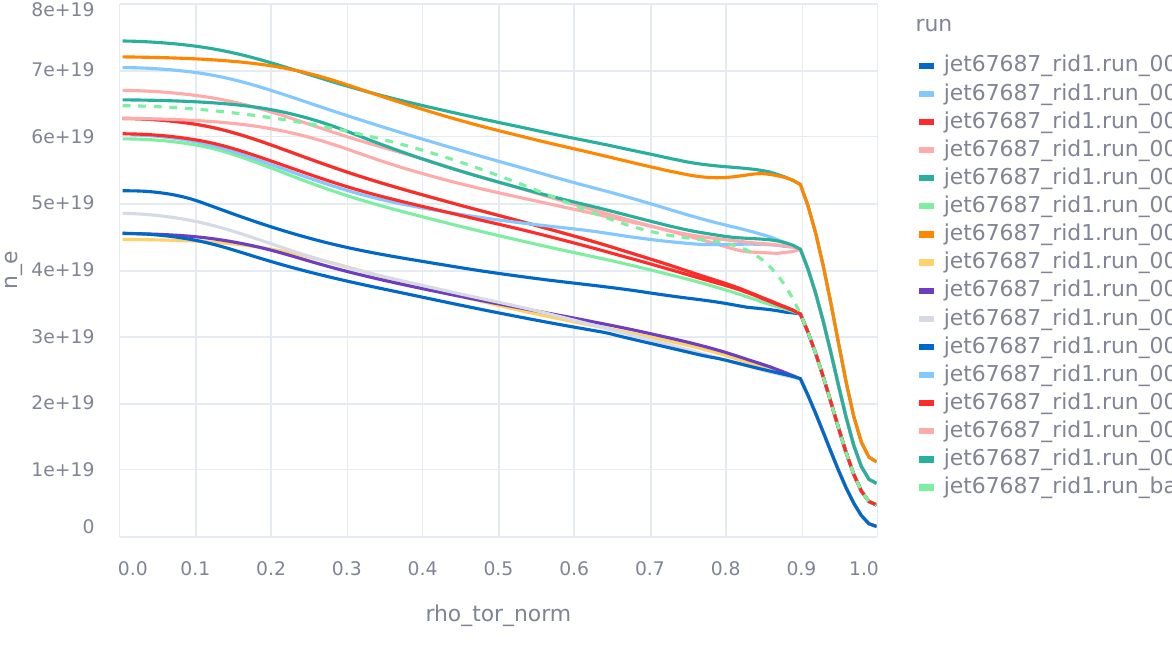}
    \end{subfigure}%
    \hspace{0.02\textwidth}%
    \begin{subfigure}[ht!]{0.45\textwidth}
    \includegraphics[width=\textwidth,trim=0 0 130 0,clip]{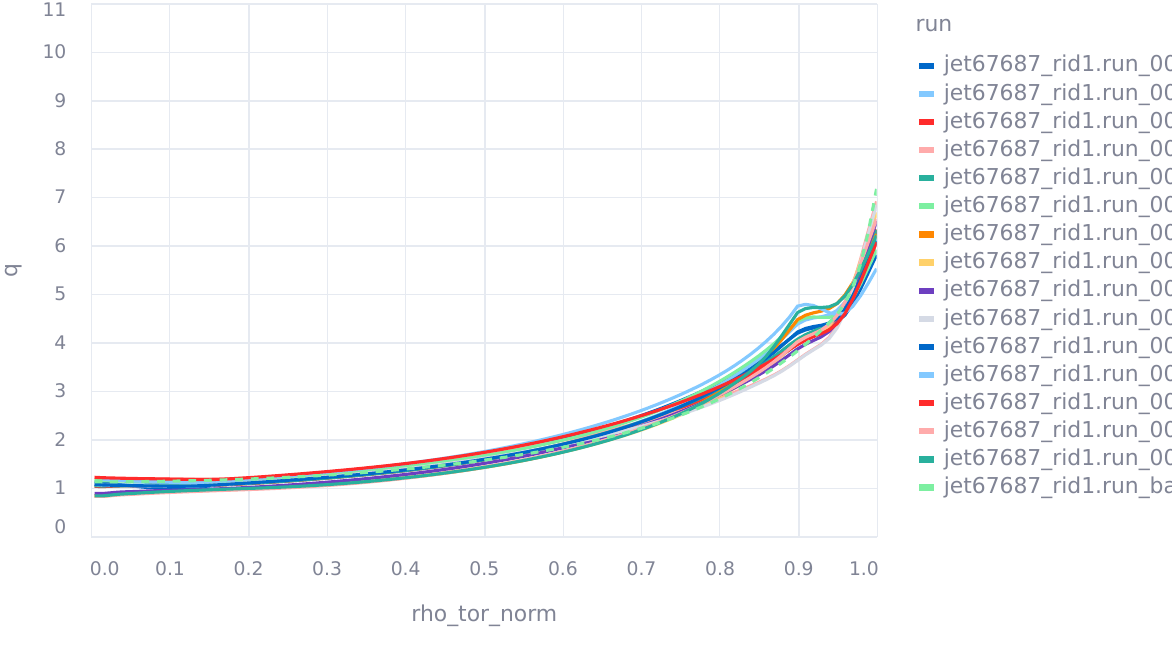}
    \end{subfigure}
    \caption{Comparison of the simulation output profiles (solid) against the input profiles (dashed) for the electron temperature, $T_e$ (top left), the average ion temperature, $T_i$ (top right), the electron density, $n_e$ (bottom left), and the safety factor, $q$ (bottom right), using a Latin-hypercube sampling on the initial profile modification based on the combined measurement and fit errors of the input profiles.}
    \label{fig:latin-hypercube-sampling-output}
\end{figure}

Figure~\ref{fig:latin-hypercube-sampling-output} shows the results of this random sampling of the input variables on the steady-state output of the JINTRAC simulation, as compared to the input condition. By looking at the collection of simulations as a whole, it appears that the variation of the output profiles for this given scenario is predominantly determined by the value of the boundary condition. However, we also saw that there are particular simulations where the output profile shapes are notably different from the majority. While the reduction of this collection of simulation outputs into a mean profile and standard deviation can be useful in succinctly describing the results of the sensitivity study, large enough deviations in the input profiles can end up in situations where the nonlinearity of the plasma transport system leads to a different solution space. This effectively means that this reduction has a slight risk of oversimplifying the results if blindly applied.

From this set of 25 randomly sampled variations, 10 of them did not successfully complete until the requested plasma time had elapsed. 5 of these crashed upon initialization due to the fact that the input uncertainty is symmetrical and the sampler chose an electron density, $n_e$, variation of $-2\sigma$, yielding negative values for the input density profiles. 3 of these crashed due to errors in the initial equilibrium reconstruction step of the simulation. The remaining 2 of these crashed due to the existence of negative temperatures, identified to be likely due to the combination of the lower electron temperature of the sampled variation and the computed radiated power.

Probing deeper in the characteristics of the completed simulations, it was found that a positive gradient occurs near the internal boundary condition region, set at the pedestal top, for the electron temperature, $T_e$, output profiles of some simulations of this particular discharge. This seems to occur specifically under a combination of a lower input safety factor, $q$, profile and a higher $T_i/T_e$ ratio, as determined by the selection of input electron and ion temperature profiles. While it is not expected that these particular insights generalize to all possible tokamak plasma regimes, performing similar sensitivity studies across a large experimental dataset can facilitate the discovery of similar trends which are robust over a wide range of plasmas. It should also be noted that while the QLKNN-jetexp-15D surrogate model has been trained using data from the JET experimental database, it has not been closely examined whether or not the simulation trajectory leaves the training domain as the simulation progresses. Therefore, it is uncertain whether the anomalies in plasma shapes are due to physically consistent outcomes of out-of-distribution combinations of perturbed parameters in the UQ scan, or due to QLKNN-jetexp-15D surrogate mode inputs being out of distribution for these cases. Dedicated investigation of the provenance of these anomalies is outside the scope of this paper.

In the light of this rudimentary sensitivity analysis, the execution and analysis of simulation swarms such as this allows common trends to be analyzed in terms of their robustness or dependency within a given experimental plasma scenario. Again, it is noted that the automated capabilities provided by \duqtools allows the setup of all these runs to be completed in approximately 1 minute, limited mostly by file read/write operations on the distributed computing system. While the human time gain on absolute terms is less in this example, it was only performed on a single time window and will accumulate to a significant amount across a large-scale exercise. In addition, the flexibility of the framework to include more sparse sampling methods and more complex user-defined operations via the YAML interface is foreseen to further decrease the time to generate insightful simulation outputs.


\section{Discussion}

Due to the acceleration of JINTRAC simulation setup via \duqtools, a preliminary large-scale validation exercise was carried out on a large number of JET experimental steady-state plasma scenarios, while both properly modifying the required discharge metadata and guaranteeing a consistent set of physics / numerical settings. This feature has already proven itself useful by illuminating potential issues in the data processing pipeline and the validity of specific simulation settings. Additionally, the capability of the tool to facilitate the multivariate exploration of input sensitivities can help analyze the robustness of certain plasma regimes and estimate the measurement tolerances needed to try avoiding those nonlinear effects in a given plasma scenario. While the prerequisite JET data has been accumulated and mentioned as part of the demonstration of this tool, the detailed analysis of those simulations and their sensitivity studies is outside the scope of this paper. Nonetheless, the basic examples and insights of these capabilities outlined here can be extended to the entire JET dataset and to data from other tokamak devices, and indeed it is recommended that this be done as well as this tool be used for facilitate further work along those directions.

To make the case for further adoption, coupling the release of this tool with an ecosystem of simulation templates would enable increased automation of executing a JINTRAC simulation on generic experimental plasma scenarios. From this platform, additional efforts can be made to build confidence in the predictive capability of the simulation suite. Due to the generality built into \duqtools, this pipeline can in principle be extended to any simulation software connected to the \imas infrastructure, even those outside the domain of plasma transport physics.

\section{Conclusions}

In this work, we describe the development of \duqtools, a tool for automatically setting up tokamak fusion plasma reactor simulations based on the \imas infrastructure. The applications show that \duqtools is a flexible tool for setting up various workflows for large scale validation with little coding. \duqtools supports sharing of workflows using template configurations. 

We have outlined the steps required for fully automated workflows, from generating a single confidence interval to large scale validation for thousands of datasets. We hope that this work can act as a blueprint to base future experiments on. Our goal is to enable canonical uncertainty quantification, making it easier for workers in the field of fusion research to generate confidence intervals for their data using a standard set of rules.

\ack

This work was supported by the Netherlands eScience Center under grant number NLESC.OEC.2021.030.
This work has been carried out within the framework of the EUROfusion Consortium, funded by the European Union via the Euratom Research and Training Programme (Grant Agreement No 101052200—EUROfusion).
This work was carried out in collaboration with the TSVV-11 team.

\clearpage
\newcommand{\newblock}{}
\bibliography{manuscript}

\end{document}